\documentclass{article}
\usepackage{graphicx,amsmath}

\newcommand{\bea}{\begin{eqnarray}}
\newcommand{\eea}{\end{eqnarray}}
\newcommand{\be}{\begin{equation}}
\newcommand{\ee}{\end{equation}}

\def\be{\begin{eqnarray}}
\def\ee{\end{eqnarray}}
\def\bd{\begin{displaymath}}
\def\ed{\end{displaymath}}

\begin{document}
\title{Nuclear structure and decay properties of even-even nuclei in $Z=70-80$ drip-line region}


\author{S. Mahapatro$^1$\footnote{Email:narayan@iopb.res.in}, C. Lahiri$^{2*}$\footnote{clahiri@iopb.res.in}, 
Bharat Kumar$^2$, R. N. Mishra$^1$ and S. K. Patra$^2$\\
$^1$Department of Physics, Ravenshaw University, Cuttack-753003, India.\\
$^2$Institute of Physics, Sachivalaya Marg, Bhubaneswar 751005, India.}

\maketitle

\begin{abstract}
We study nuclear structure properties for various isotopes of Ytterbium
(Yb), Hafnium(Hf), Tungsten(W), Osmium(Os), Platinum(Pt) and Mercury(Hg) 
in $Z = 70 - 80$ drip-line region  starting from $N =80$ to $N=170$  
within the formalism of relativistic mean field (RMF) theory. The pairing correlation is taken care by using BCS approach. 
We compared our 
results with Finite Range Droplet Model(FRDM) and experimental data and
found that the calculated results are in good agreement. Neutron shell closure are obtained at $N=82$ and $126$ in this region. 
We have also studied probable decay mechanisms of these elements.

\end{abstract}

PACS Number(s): 21.10.Dr,21.10.Pc,21.10.Tg,23.40.-s,23.60.+e


\section{Introduction}\label{introduction}
Recent developments of Radioactive Ion Beam (RIB) facilities has encouraged various experimental as well as theoretical 
developments  for nuclei far from the $\beta$ stability line. 
Physics of nuclear structure and decay mechanism in neutron rich nuclei is a very popular field of study
 in present days~\cite{hama, exrich}. Recently the
study of deformed nuclear shapes is also a very interesting topic
and a number of excellent review articles exist
~\cite{nolan88, flocard90, baktash95, bak97} in this context.
In literature, after the early calculations, which have 
predicted the existence of the superdeformed shape (see
Ref.[2] for the review), its first experimental discovery
took place in 1985 ~\cite{twin86}. Till then numerous groups have developed
sophisticated techniques to perform calculations
allowing the description of rotational bands
for very elongated shapes. The phenomenon of
superdeformation constitutes a spectacular 
manifestation of the mean-field properties of
nuclei.

The change of shape of nuclei and  other properties
of nuclear structure with neutron number has attracted
much theoretical and experimental attention for many years
~\cite{quentin78, baranger68, alkha89, thi81, lee74}.
The conditions needed for the observation of shapes
coexisting in such nuclei have been a matter of 
challenge and theoretically calculated by Patra and Panda~\cite{panda93} in 1993.
The change of shapes of Pt isotopes was  
studied by Sharma and Ring~\cite{sharma92}.
The shapes of nuclei such as Os and Pt have also been 
studied by using the Hartree-Fock Bogoliubov
(HFB) formalism within the  non-relativistic
framework~\cite{quentin78}. Recently B. Kumar {\it et al.}~\cite{bkumar} studied the shape co-existence in Zr isotopes.

The nuclei with $Z=70-80$ consists of both Rare Earth Metals(Yb) as well as the Transition metals(Hf, W, Os, Pt, Hg). They manifest a 
lot of interesting phenomena like shape co-existence, change of shape
in an isotopic chain etc~\cite{zn,prcr}. In this region of the atomic
number, variety of shapes such as prolate, oblate and
spherical configuration of nuclei appears in ground state\cite{lala_all}.

Many theoretical as well as experimental papers have focused 
their interest on the shape transition~\cite{dabki79}, shape isomerism and the 
observation of superdeformed bands~\cite{girod93} in neutron-deficient Hg 
isotopes. 

The aim of this manuscript is to explore the neutron rich side of the concerned mass region with RMF formalism. 
Mainly the structure informations like deformation, two neutron separation energy, single particle energy levels
etc. have been extracted using NL3 force parameter.  

Further, $\alpha$ decay has been remained a very powerful tool to study the nuclear structure since its discovery
by Becquerel in 1896. We also find other exotic decay modes like $\beta$ decay, spontaneous fission,cluster-decay etc. Therefore,
it will be interesting to see the preferred decay modes of these neutron-rich even-even nuclei.



The paper is organized as follows.
In section ~\ref{rmf}, we have
given a brief outline about the relativistic mean field (RMF)
formalism.
The effects of pairing for open shell nuclei, included
in our calculations, are also discussed in this section.
Our results are discussed in section~\ref{result}.
The reaction Q values for $\alpha$ and $\beta$ decay and their corresponding half lives are
given in section ~\ref{decay}.
A concluding remark is given in section ~\ref{remark}.

\section{Theoretical Framework for Relativistic Mean Field Model}\label{rmf}
The relativistic mean field (RMF) model 
~\cite{gamb90, patra91, wal74, sero86, horo81, bogu77, price87, niksic11} is  
a well applied technique in recent years and have been applied to finite nuclei 
and infinite nuclear matter. In the present work,
we have taken the RMF Lagrangian ~\cite {gamb90} 
with NL3 parameter set ~\cite{lala97} in our study. 
This force parameter set is successful 
in both $\beta$-stable and drip-line nuclei. 
The Lagrangian contained the term of interaction between meson and nucleon 
and also self-interaction of isoscalar-scalar {\it sigma} meson. 
The other mesons are isoscalar-vector {\it omega} and isovector vector 
{\it rho} mesons. The photon field $A_{\mu}$ is included to take care of 
Coulombic interaction of protons.  
The pairing correlation is taken care by using BCS approach ~\cite{gamb90}.
We start with the relativistic Lagrangian density for a nucleon-meson 
many-body system,
\begin{eqnarray}
{\cal L}&=&\overline{\psi_{i}}\{i\gamma^{\mu}
\partial_{\mu}-M\}\psi_{i}
+{\frac12}\partial^{\mu}\sigma\partial_{\mu}\sigma
-{\frac12}m_{\sigma}^{2}\sigma^{2}\nonumber\\
&& -{\frac13}g_{2}\sigma^{3} -{\frac14}g_{3}\sigma^{4}
-g_{s}\overline{\psi_{i}}\psi_{i}\sigma-{\frac14}\Omega^{\mu\nu}
\Omega_{\mu\nu}\nonumber\\
&&+{\frac12}m_{w}^{2}V^{\mu}V_{\mu}
+{\frac14}c_{3}(V_{\mu}V^{\mu})^{2} -g_{w}\overline\psi_{i}
\gamma^{\mu}\psi_{i}
V_{\mu}\nonumber\\
&&-{\frac14}\vec{B}^{\mu\nu}.\vec{B}_{\mu\nu}+{\frac12}m_{\rho}^{2}{\vec
R^{\mu}} .{\vec{R}_{\mu}}
-g_{\rho}\overline\psi_{i}\gamma^{\mu}\vec{\tau}\psi_{i}.\vec
{R^{\mu}}\nonumber\\
&&-{\frac14}F^{\mu\nu}F_{\mu\nu}-e\overline\psi_{i}
\gamma^{\mu}\frac{\left(1-\tau_{3i}\right)}{2}\psi_{i}A_{\mu} . 
\end{eqnarray}
Here sigma meson field is denoted by $\sigma$, omega meson field by $V_{\mu}$ 
and rho meson field is denoted by ${\rho}_{\mu}$. In the equations, $A^{\mu}$ denotes the
electromagnetic field, which couples to the protons. ${\psi}$
are the Dirac spinors for the nucleons, whose third components of isospin is
$\tau_{3}$ and $g_{s}$, $g_2$, $g_3$, $g_{\omega}$,$c_3$, 
$g_{\rho}$ are the coupling constants.
A definite set of coupled equations are
obtained from the above Lagrangian and solved  self-consistently
in an axially deformed harmonic oscillator basis.
The total energy of the system is given by the expression,

\begin{equation}
E_{total}=E_{part} + E_{\sigma} +  E_{\omega} + E_{\rho} + 
E_{c} + E_{pair} + E_{c.m.},
\end{equation}
where $E_{part}$ is the sum of the single-particle energies
of the nucleons and $E_{\sigma}$, $E_{\omega}$, $E_{\rho}$,
$E_{c}$, $E_{pair}$ are the contributions of the meson field,
the coulomb field and the pairing energy respectively. The center of mass (c.m.) motion energy correction is
estimated by the harmonic
oscillator formula $E_{c.m.} = \frac{3}{4}(41A^{-1/3})$ MeV, where $A$ is the
mass number of the nucleus. The total quadrupole deformation parameter 
$\beta_{2}$ and hexadecoupole parameter $\beta_{4}$ of the nucleus can be obtained from the quadrupole moment $Q$ and hexadecoupole moment $H$, respectively
using the relations 
\begin{equation}
\beta_{2}=\sqrt{\frac{5\pi}9}\frac {Q}{AR^{2}} ~~~\&   ~~\beta_{4}=\frac{4\pi}{3}\frac {H}{AR^{4}},
\end{equation}
where R is the nuclear radius.
The root mean square (rms) matter radius is given as 
\begin{equation}
\langle r_m^2 \rangle = {1\over{A}}\int\rho(r_{\perp},z).
r^2d\tau 
\end{equation}
Here $\rho(r_{\perp},z)$ is the deformed density. The total binding 
energy and other observables are also obtained by using the standard 
relations, given in ~\cite{gamb90, patra91}. 
\subsection{Pairing Correlations in RMF formalism}\label{sec:bcs}
Pairing correlation is playing very crucial role in open shell nuclei. In our 
calculation we are using the Bardeen-Cooper-Schrieffer (BCS) pairing for 
determining the bulk properties like binding energy (BE), quadrupole 
deformation parameter and nuclear radii. 
The pairing energy can be given as:
\begin{equation}
E_{pair}=-G\left[\sum_{i>0}u_{i}v_{i}\right]^2,
\end{equation}
where G is pairing force constant and $v_i^2$, $u_i^2=1-v_i^2$ are the
occupation probabilities respectively ~\cite{patra93, gamb90, pres82}. The simple
form of BCS equation can be derived from the variational method with respect 
to the occupation number $v_i^2$ and is given by:
\begin{equation}\label{eq:BCS} 
2\epsilon_iu_iv_i-\triangle(u_i^2-v_i^2)=0,
\end{equation} 
using $\triangle=G\sum_{i>0}u_{i}v_{i}$.

The occupation number is defined as:
\begin{equation}
n_i=v_i^2=\frac{1}{2}\left[1-\frac{\epsilon_i-\lambda}{\sqrt{(\epsilon_i-\lambda)^2+\triangle^2}}\right].
\end{equation}

In this calculation we are dealing with the constant gap formalism for 
proton and neutron in a region far from beta stability line. 
These constant gap equation for proton and neutron is taken from Ref. 
~\cite{madland81,moller88} which is given as:

\begin{equation}
\triangle_p =RB_s e^{sI-tI^2}/Z^{1/3}
\end{equation}
and
\begin{equation}
\triangle_n =RB_s e^{-sI-tI^2}/A^{1/3},
\end{equation} 
with $R$=5.72, $s$=0.118, $t$= 8.12, $B_s$=1, and $I = (N-Z)/(N+Z)$.
In our present calculation, we have taken the constant pairing gap for all 
states $\mid\alpha>=\mid nljm>$ near the Fermi surface for the shake of 
simplicity. Similar approach was used in the references~\cite{new1,new2,new3}.

\begin{figure}[ht]
\includegraphics[scale=0.3]{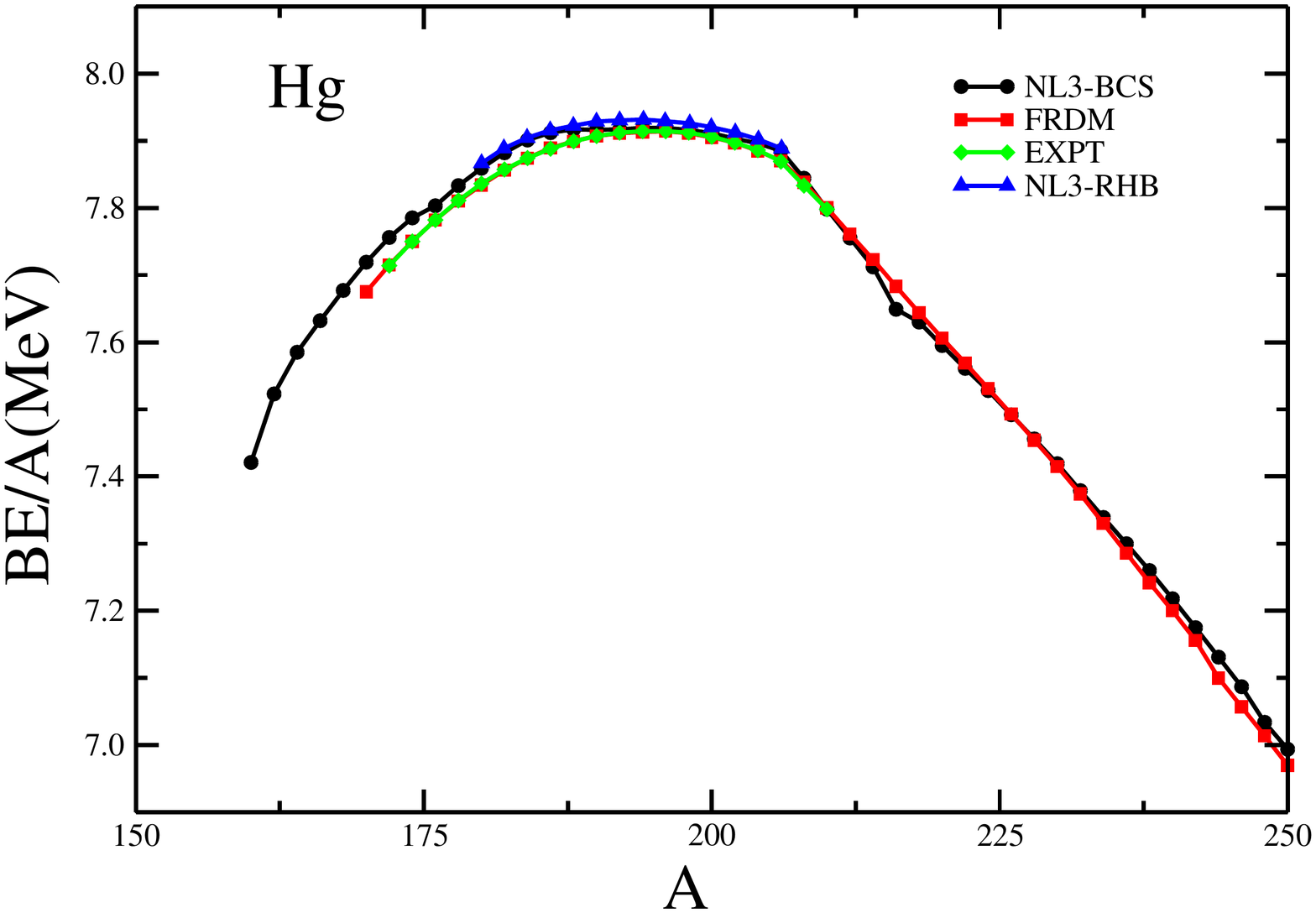}
\caption{(Color online) The binding energy per particle (BE/A) obtained from
RMF(NL3-BCS) (circle) compared with RHB (triangle)~\cite{rhb}
FRDM (square)~\cite{moll97} and
experimental (diamond)~\cite{audi12} data for Hg isotopes.}
\label{be}
\end{figure}

As we know, if we go near the neutron drip line, then coupling to the 
continuum become important ~\cite{doba84,doba96,meng1,meng2}. In this case  
the Relativistic Hartree-Bogoliubov (RHB) approach~\cite{new4,new5} has been proved to be a more accurate 
formalism for this region. However in case of RMF-BCS formalism also, the pairing constants along with NL3 parameters 
are adjusted so that it can reproduce the bulk properties of a nucleus near the drip line region. For example, 
in Fig. \ref{be} the total binding energy/nucleon(BE/A) for Hg isotopes 
from our calculation are compared with available NL3-RHB data\cite{rhb} along with existing FRDM~\cite{moll97} and experimental
results~\cite{audi12}. The other components of our present calculation like the selection of basis space
is described in the next section.
It is evident from the figure that the binding energy from our calculation is in a nice agreement with FRDM and 
experimental results in neutron rich region. Further, our present result agrees acceptably well with the available RHB data.
By using BCS pairing correlation model,
it has been shown earlier  by different groups ~\cite{patra01,werner94,werner96,lala01,lala99} 
that the results from relativistic mean field BCS (RMF-BCS)
approach is almost similar with the RHB formalism. 
However due to the lack of NL3-RHB data in neutron rich region for the concerned isotopes, we are unable to compare 
their individual merits, but the above figure and references suggest that the present formalism can be extended
in neutron rich region.

\subsection{Selection of Basis Space}\label{sec:basis}
\begin{figure}[ht]
\includegraphics[scale=0.3]{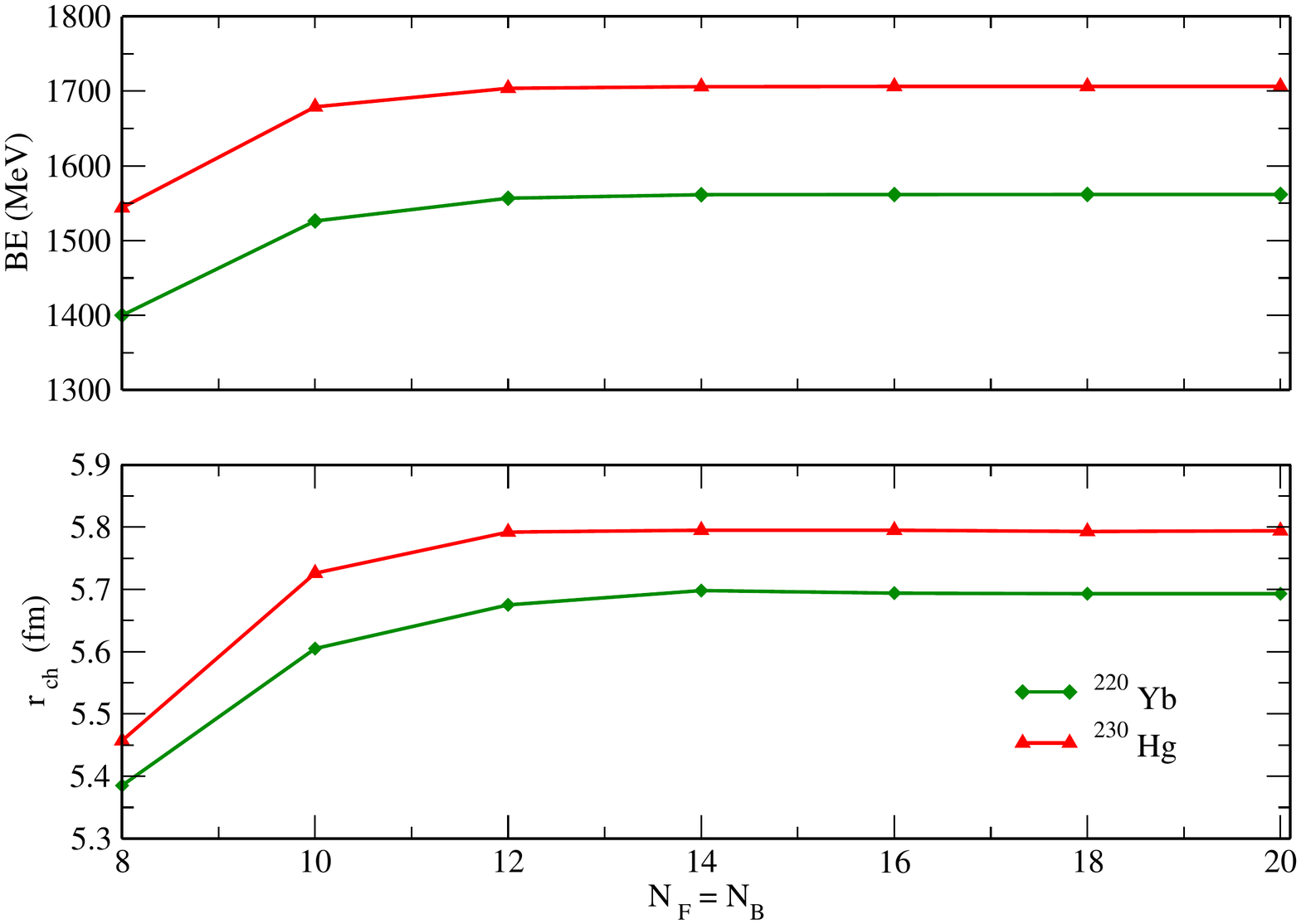}
\caption{ (Color online)  The variation of calculated binding energy (BE), charge
radii (r$_c$) with increasing number of Bosonic and Fermionic basis.}
\label{basis}
\end{figure}

In order to choose the proper basis, we calculate the physical observables like 
binding energy, root mean square(rms) radii and quadrupole deformation 
parameter($\beta_2$).  To select optimal values
for $N_F$ and $N_B$, we select $^{220}$Yb and $^{230}$Hg as  test case and increase the
basis quanta from 8 to 20 step by step in Fig. \ref{basis}. We find that
these physical observables varies with changing the bosonic and 
fermionic oscillator quanta  $N_F$=$N_B$= $8$ to $12$ but become almost constant 
after $N_F$=$N_B$= $14$. But our calculation is extended far beyond the stability line which 
demands a comparatively large basis space for calculation. Therefore for greater accuracy, we have chosen $N_F$=$N_B$= $20$ for all the calculations. Similar types of calculation are found in Refs.\cite{patra05, bha,bkumar,gamb90,gam94}. 


\section{Calculations and Results}\label{result}
Relativistic mean field model have given very good result in $\beta$- stable 
nuclei of the nuclear landscape. In this work we are analyzing the exotic 
neutron drip line nuclei by using RMF model with well known 
NL3 ~\cite{lala97} parameter set. In 1999, Lalazissis {\it et al.} ~\cite{lala_all} analyzed ground-state properties of 1315 even-even
nuclei with $Z$ between 10 and 98 and calculated total binding energy, radius and deformations using similar formalism mostly near the 
stability region . In the present 
manuscript, we have extended the RMF-BCS formalism near the neutron drip line for even-even nuclei in $Z=70-80$ region. In Tables 1-6, the ground state binding energy, neutron, proton, charge and r.m.s. radius, quadrupole and hexadecoupole parameter of the above isotopes
are compared with FRDM ~\cite{moll97,moll95} and experimental data~\cite{audi12,raman01,angeli13}. 
In the upcoming subsections our results are described in detail.

\subsection{Quadrupole Deformation}
\begin{figure*}[ht]
\includegraphics[scale=0.5]{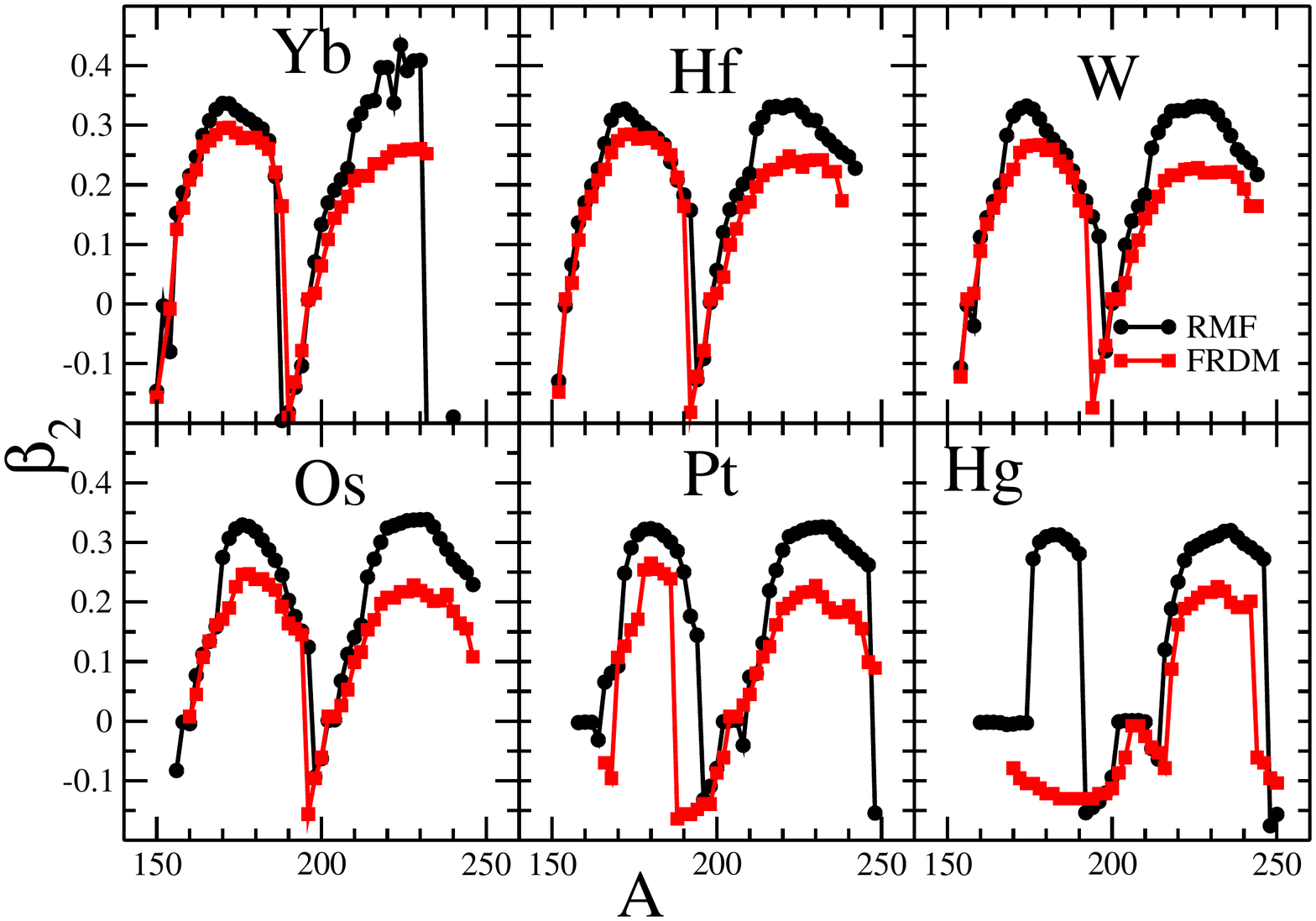}
\caption{(Color online) The quadrupole moment deformation parameter $\beta_2$
obtained from RMF(NL3)(circle) compared with the Finite Range
Droplet Model(FRDM) (square)~\cite{moll95} 
results for different isotopes of
Yb, Hf, W, Os, Pt and Hg in $Z =70-80$
drip-line region.}
\label{BETA}
\end{figure*}

The Quadrapole deformation parameter $\beta_{2}$ for the ground state
 for even-even nuclei in $Z=70-80$ region are determined within the RMF-BCS formalism near the neutron drip-line 
 as an extension of existing data\cite{lala_all}.
The parameter $\beta_{2}$  is directly connected to the shape of the nucleus. 
The ground state quadrupole deformation parameter $\beta_{2}$ is plotted in
Fig. \ref{BETA}. We have compared our RMF results with FRDM ~\cite{moll95} results.
                  
In case of Yb isotopes in Fig. \ref{BETA}, RMF $\beta_{2}$ data
coincides  with FRDM ~\cite{moll95}
values for almost all the mass numbers.
In lower mass regions, Yb isotopes are nearly spherical($A =150$).
It changes to oblate at $A =155$. The prolate stage exists from
$A =160-190$. At $A =190-200$, it changes from spherical to oblate.
At $A =200-230$ again it changes to  prolate state.
In lower and higher mass region, maximum isotopes are prolate in
shape. In the middle region, these changes to oblate shape.
Therefore one can say that, in Yb isotopes, at $A =155$ and $190$, there is a shape transition from oblate to
highly prolate . FRDM also shows the same trend in shapes.
The same trend is found in other isotopes except mercury (Hg).
Mercury shows prolate shape at $A=170-190$ whereas FRDM shows oblate shape
in the same region.

\subsection{Two-neutron separation energy}
\begin{figure*}[ht]
\includegraphics[scale=0.5]{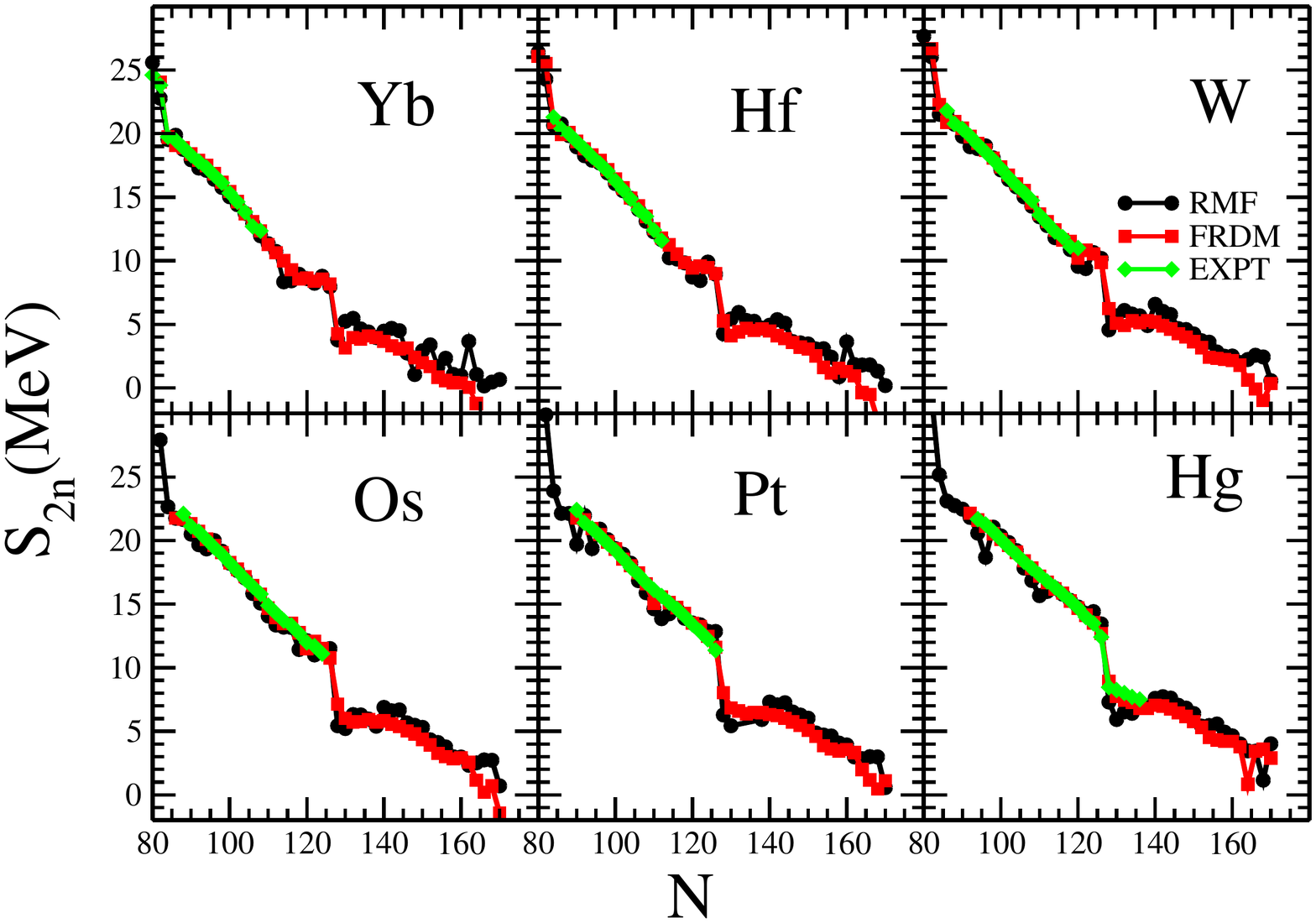}
\caption{(Color online) Two-neutron separation energy $S_{2n}$
(RMF)(circle) is compared with FRDM (square) and experimental results (diamond)
for different isotopes of $Z=70-80$ region.
}
\label{s2n}
\end{figure*}

\begin{figure*}[ht]
\includegraphics[scale=0.5]{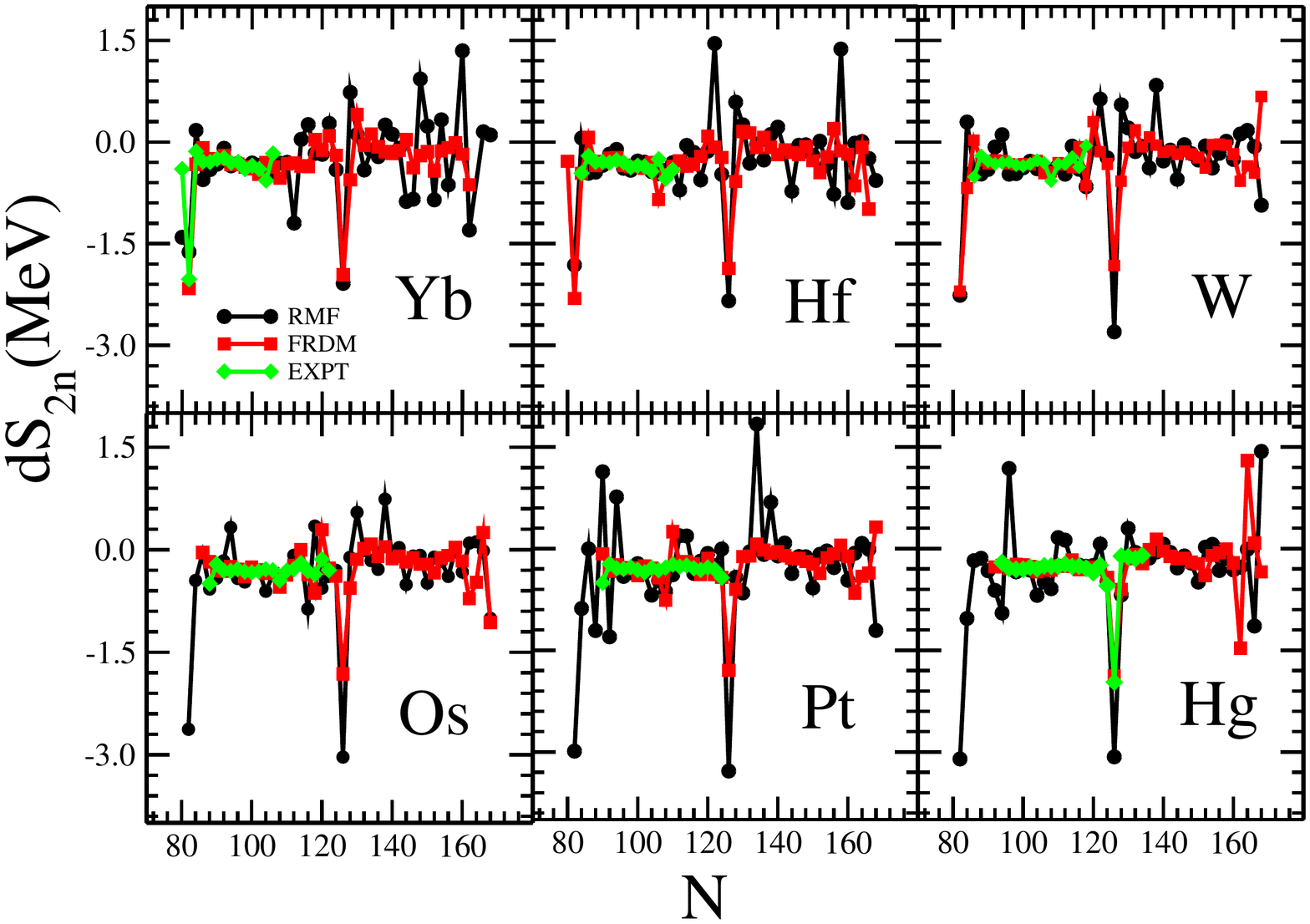}
\caption{(Color online) The differential variation of two-neutron separation energy $dS_{2n}$
(RMF)(circle) is compared with FRDM (square) and experimental results (diamond)
for different isotopes of $Z=70-80$ region.
}
\label{ds2n}
\end{figure*}

The two-neutron separation energy $S_{2n}$(N,Z) =  BE(N, Z) - BE(N-2, Z) is
shown in Fig. \ref{s2n} for Yb, Hf, W, Os, Pt and Hg isotopes. 
The  $S_{2n}$ values calculated from RMF, FRDM~\cite{moll97} and Experimental~\cite{audi12} results are compared in Fig. \ref{s2n}. We can predict the stability of these nuclei by $S_{2n}$ energy.
If $S_{2n}$ is large, it means nuclei will be stable with two-neutron
separation whereas the two neutron drip line for an isotopic chain can be identified as the nucleus having zero or slightly positive separation energy value.
Therefore, it is evident from the figure that the two neutron drip line for the $Z=70$(Yb) isotope comes at $N=166$, the same for $Z=72,74,76$ isotopes is approximately at $N=170$. In case of  $Z=80$(Hg), the two neutron drip line arrives at the neutron number $N=168$.
Furthermore, in all cases,
the $S_{2n}$ values decrease gradually with increase in neutron number.
There is a kink appeared at neutron magic number $N=126$ in every cases.
In order to visualize this scenario with a bit clarity, 
in the next subsection, we defined differential variation of two-neutron separation energy given in Fig. \ref{ds2n}. 

\subsection{Differential variation of two-neutron separation energy}
The differential variation of the two-neutron separation energy $S_{2n}$
with respect to the neutron number(N), i.e., $dS_{2n}$(Z, N) is defined as:

\begin{equation} 
dS_{2n}(Z, N) =\frac{S_{2n}(Z, N+2) - S_{2n}(Z, N)}{2}.
\end{equation}

The $dS_{2n}$(Z, N) is important factor to find
the rate of change of separation energy with respect to the 
neutron number in an isotopic chain. Here, we calculated
the $dS_{2n}$(Z, N) for NL3 parameters and compared it with FRDM and experimental
results extracted from refs.~\cite{moll97} and ~\cite{audi12} respectively. 
In general, the large,
sharp, deep fall in the $dS_{2n}$(Z, N) is observed for Yb, Hf, W, Os, Pt and Hg
isotopes at N=82 and 126 ( Fig. \ref{ds2n}). It shows the neutron shell closures at $N=82$ and $N=126$. 

\begin{figure*}[ht]
\includegraphics[scale=0.5]{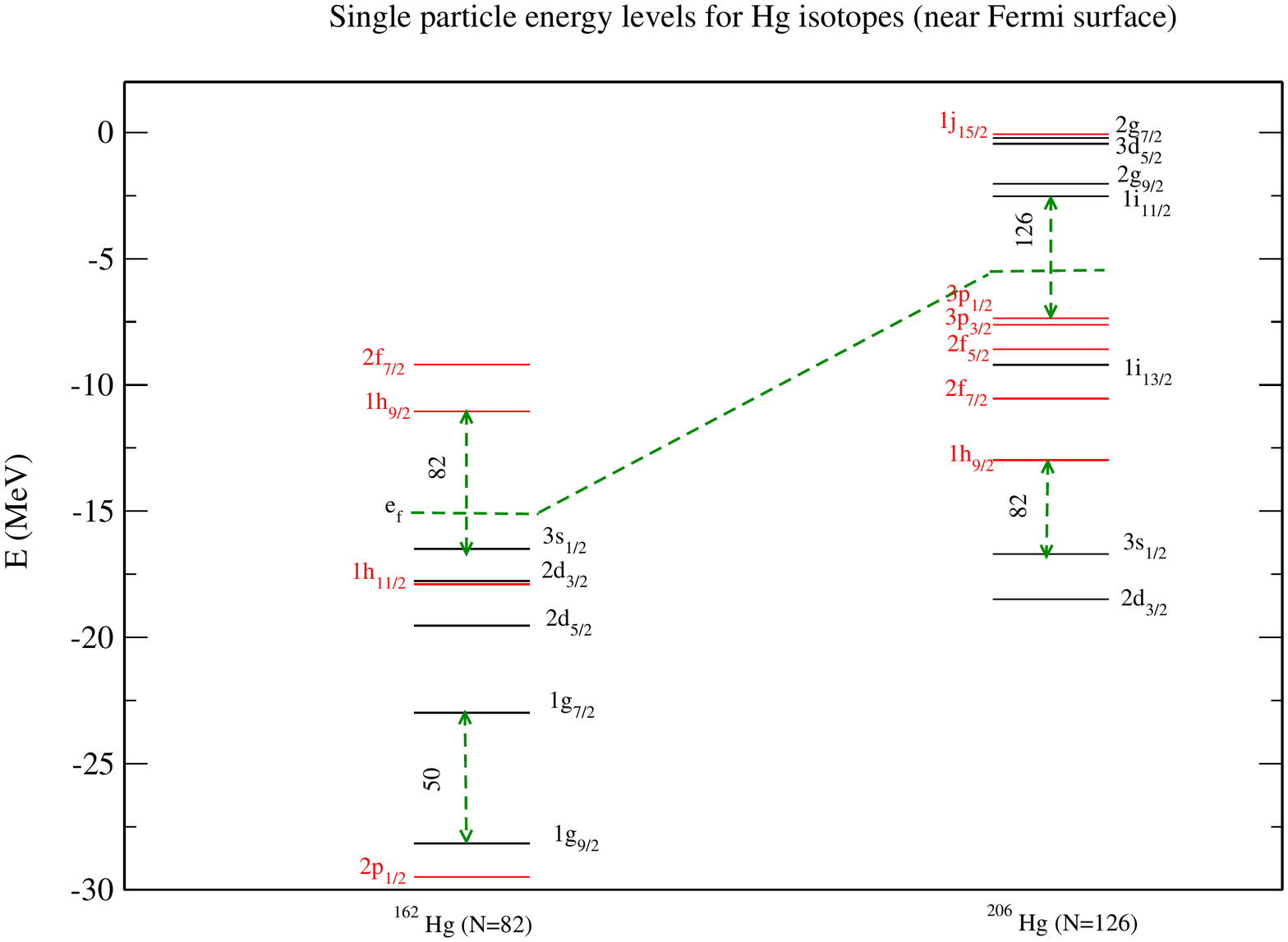}
\caption{(Color online) The single particle energy levels for $^{162}$Hg and $^{206}$Hg isotopes from RMF model with NL3 parameter set. The dashed-line indicate the fermi surface $e_f$. 
}
\label{single}
\end{figure*}

\subsection{Single Particle energy levels }
In Fig. \ref{single} we plotted single particle energies for neutron levels near the Fermi surface for Hg isotopes ($Z=80$)
at neutron shell closures ($N=82 ~\& ~126 $). We have used deformed configuration to calculate these levels. From Table \ref{Hg}, it is evident that $^{162}$Hg and $^{206}$Hg are nearly spherical and therefore the single particle levels are almost identical to that of the spherical configuration. For example, in $^{162}$Hg, the level g$_{7/2}$ actually consists of four overlapped energy states having spins ${7/2}^{+},{5/2}^{+},{3/2}^{+} ~\& ~{1/2}^{+}$.
In both cases, we found some large gaps at $N=50,82,126$ shell closures. Here, the green 
dotted line shows the approximate fermi level.
In case of Yb, Hf, W, Os and Pt also the gaps are visible at magic shell closures. However, they are not depicted in this manuscript.
\section {Decay Modes}\label{decay}
In this section we discuss probable decay modes of these neutron-rich nuclei.
\subsection{Alpha decay half life}
The $Q_{\alpha}$ energy is obtained from the relation \cite{patra97}:
$Q_{\alpha}(N, Z)$ = $BE (N, Z)$ - $BE (N - 2, Z - 2)$ - $BE (2, 2).$
Here, $BE(N, Z)$ is the binding energy of the parent nucleus with neutron
number N and proton number Z, $BE (2, 2)$ is the binding energy of the
$\alpha$-particle ($^4$He), i.e., 28.296 MeV, and $BE (N- 2, Z - 2)$ is
the binding energy of the daughter nucleus after the emission of an
$\alpha$-particle.

The expression for the $\alpha$-decay half life
from Viola and Seaborg \cite{viola66} is given by:
\begin{equation}
log_{10}T_{1/2}^{\alpha}(s) = \frac{aZ - b}{\sqrt Q_\alpha}
-(cZ+ d) + h_{log},
\end{equation}
with $Z$ as the number of proton for the parent nucleus and the constants
${a}$, ${b}$, ${c}$ and ${d}$, are from Sobiczewski et al. \cite{sobi89}.
The value of these constants are ${a}$ = 1.66175, ${b}$ = 8.5166, ${c}$ 
= 0.20228, ${d}$ = 33.9069 and the quantity $h_{log}$ accounts for the 
hindrances associated with the odd nucleon as, 
\begin{eqnarray}
h_{log}   & = & 0        for ~~\mbox{Z} ~~\mbox{even} ~~\mbox{and} ~~\mbox{N}
~~\mbox{even}~~\nonumber \\
            & = & 0. 772   for ~~\mbox{Z} ~~\mbox{odd} ~~\mbox{and} ~~\mbox{N}
~~\mbox{even} \nonumber\\
            & = & 1. 066   for ~~\mbox{Z} ~~\mbox{even} ~~\mbox{and} ~~\mbox{N}
~~\mbox{odd} \nonumber\\
            & = & 1. 114   for ~~\mbox{Z} ~~\mbox{odd}  ~~\mbox{and} ~~\mbox{N}
~~\mbox{odd}.
\end{eqnarray}

\begin{figure*}[ht]
\includegraphics[scale=0.5]{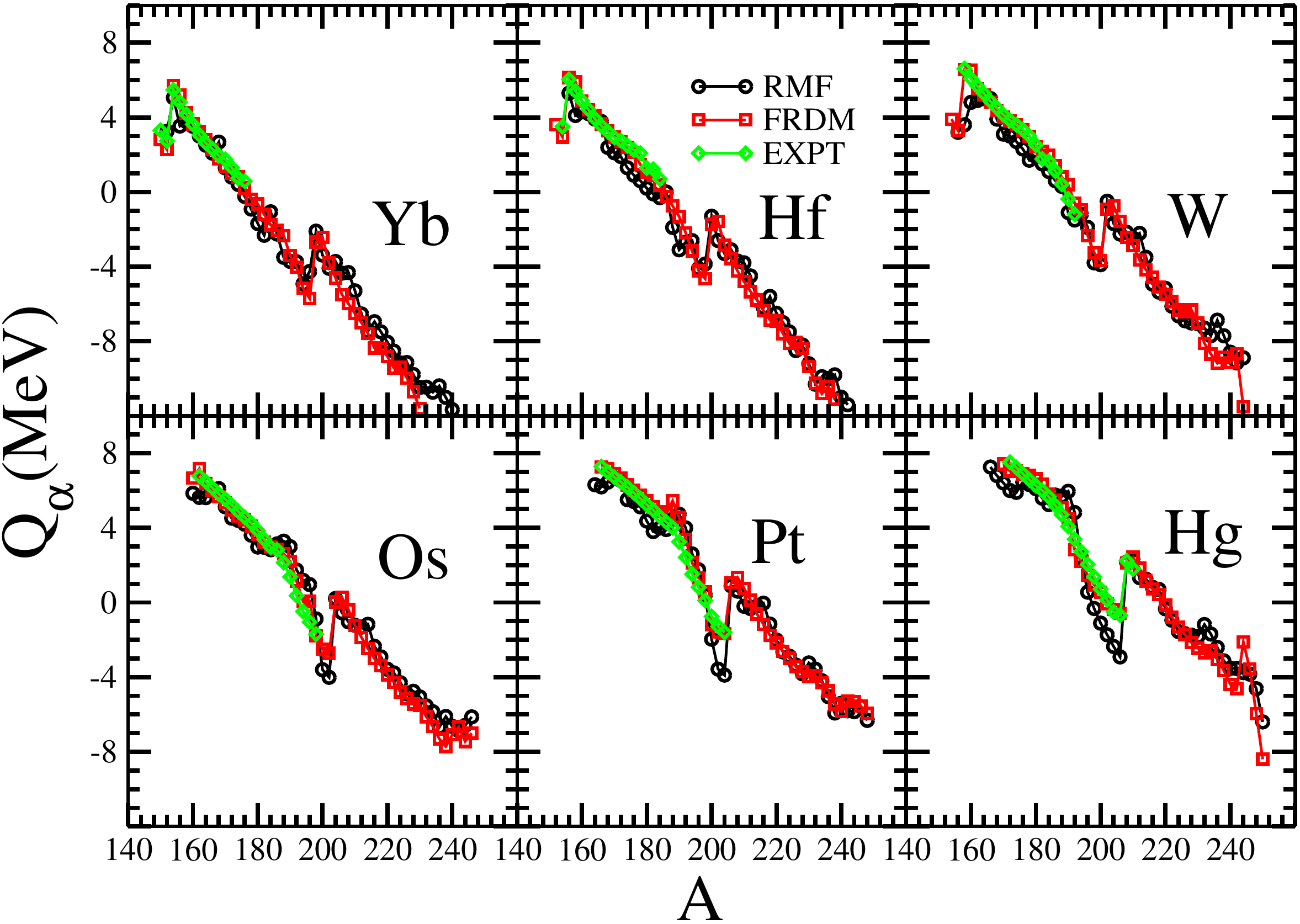}
\caption{(Color online) The $Q_{\alpha}$ energy obtained from RMF(NL3)(circle) compared with the FRDM(square) and experimental(diamond) results for different
isotopes of $Z=70-80$ region.}
\label{Qalph}
\end{figure*}

\begin{figure*}[ht]
\includegraphics[scale=0.5]{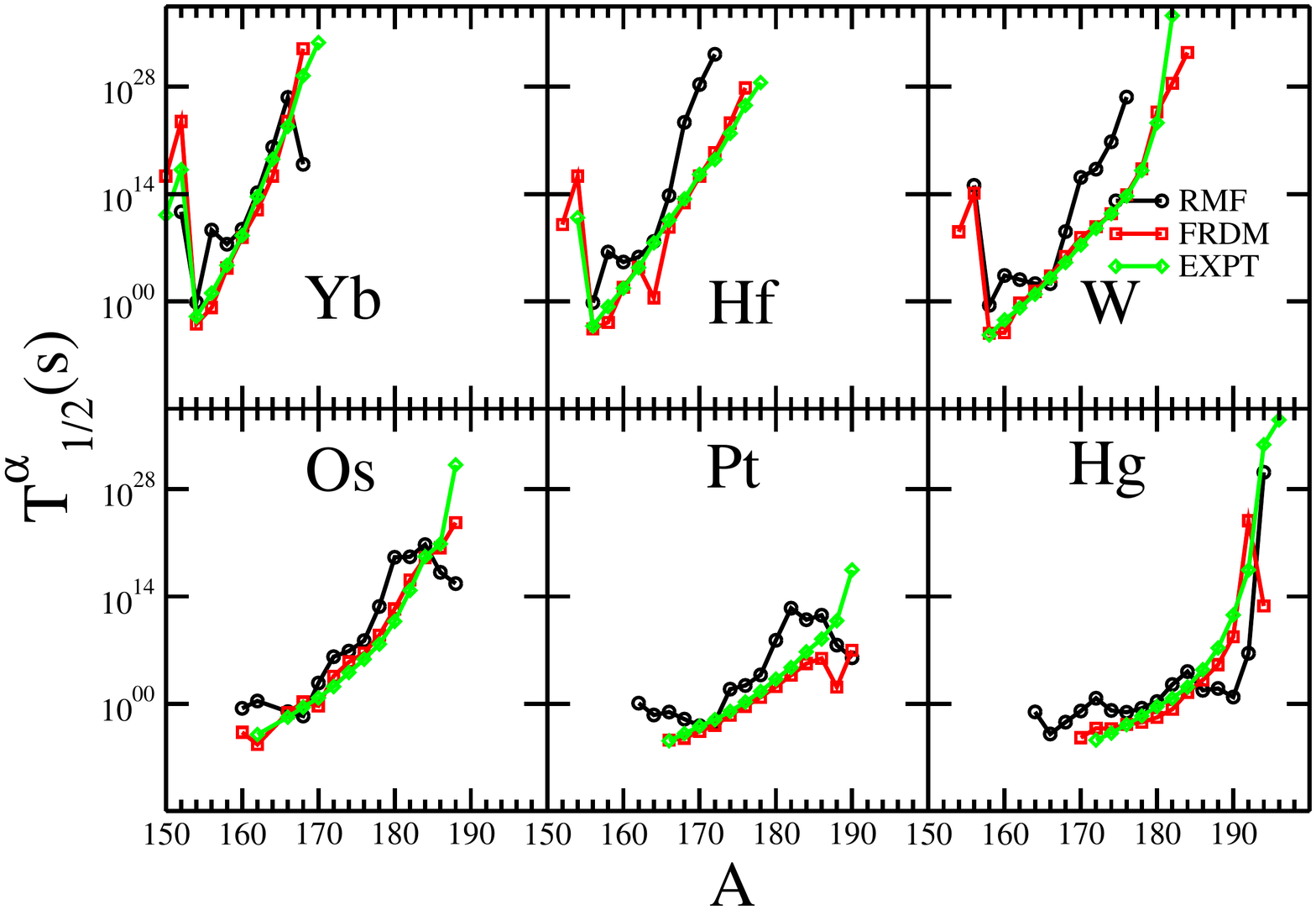}
\caption{(Color online) The $T_{1/2}^{\alpha}$   obtained from RMF(NL3)(circle) 
are compared with FRDM(square) and experimental(diamond) results
in $Z = 70-80$ region.}
\label{time}
\end{figure*}

We evaluate the BE by using RMF formalism and from these, we estimated
the $Q_{\alpha}$ for all the isotopes of $Z=70-80$  region. We have calculated
the half-life time $T_{1/2}^{\alpha}$ by using the above formulae.
The comparison of $Q_{\alpha}$ and $T_{1/2}^{\alpha}$ are shown in Fig. \ref{Qalph} and
Fig. \ref{time} respectively. From Fig. \ref{time}, it is evident that the alpha decay half life goes on increasing with large neutron numbers and 
follows similar trend for RMF, FRDM and experimental cases. In another word, we can say that, these neutron rich nuclei 
hardly exhibit $\alpha$-decay and therefore the concerned decay channel is almost forbidden in this region.   

In order to check the the usual decay mechanism of these neutron-rich samples, 
in the next subsection we studied the $\beta$ decay half-life.  

\subsection{Beta decay half life}
As discussed in the previous subsection, we expect $\beta$ decay to be the prominent mode in case of neutron-rich nuclei. In order to calculate $\beta$ decay half life, we have
used the empirical formula of Fiset and Nix\cite{fiset}. However the formula holds good mainly for
the superheavy region, but in this work we  have checked its credibility in $Z=70-80$ region. 
The formula for $\beta$ decay half life is defined as:

\begin{equation}
T_{1/2}^{\beta}=(540\times10^{5})\frac{{m_e}^5}{\rho_{d.s.}({W_\beta}^6-{m_e}^6)}.
\end{equation}

\begin{figure*}[ht]
\includegraphics[scale=0.5]{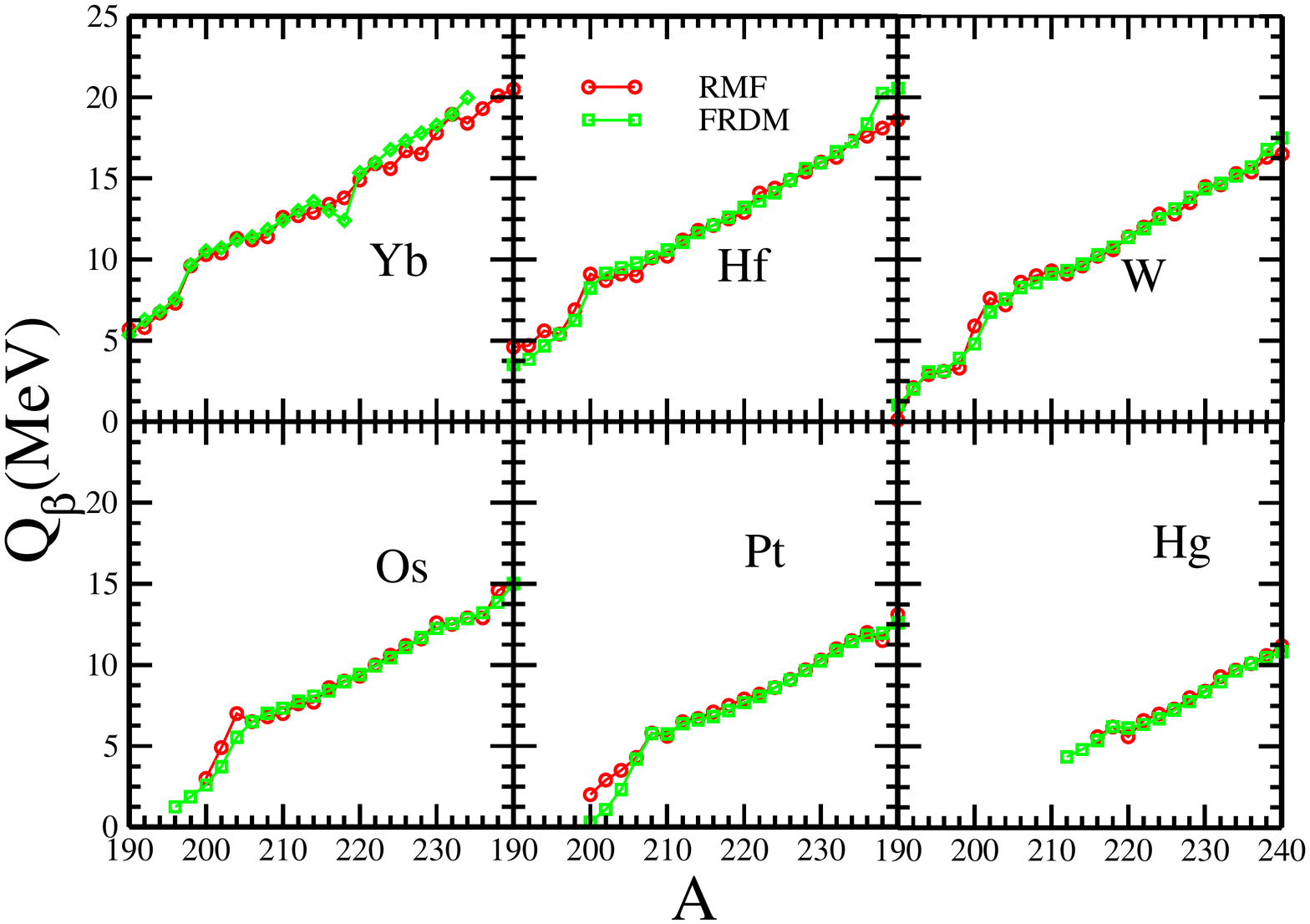}
\caption{(Color online) The $Q_{\beta}$ energy obtained from RMF(NL3)(circle) are 
compared with the FRDM(square)  data for different isotopes in $Z=70-80$ region.}
\label{Qbet}
\end{figure*}

\begin{figure*}[ht]
\includegraphics[scale=0.5]{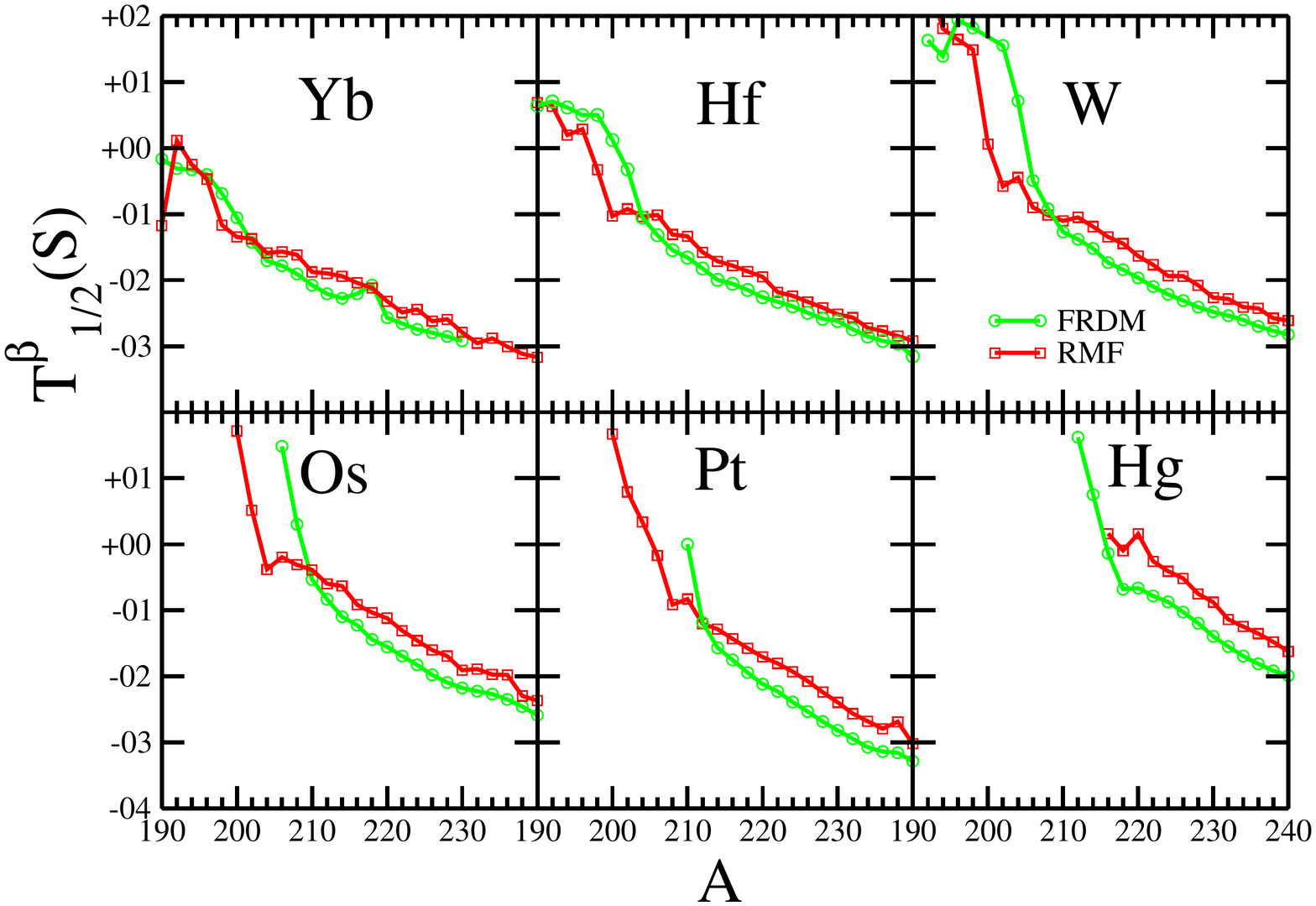}
\caption{(Color online) The $T_{1/2}^{\beta}$   obtained from RMF(NL3)(circle) 
are compared with FRDM(square) data 
in $Z = 70-80$ region.}
\label{tbet}
\end{figure*}

In the expression, $Q_\beta = BE(Z + 1, A) - BE(Z, A) 
$  is the $\beta$ decay
Q value for the sample $(Z,A)$ which decays to the daughter $(Z+1, A)$ and $W_\beta = Q_\beta + {m_e}^ 2$. 
Here, $\rho_{d.s.}$ is the average density of states
in the daughter nucleus ($e^{−A/290} \times $ number of states within 1
MeV of ground state). The approach has been applied in Th an U isotopes\cite{bha} recently.  It is important to note that all the daughter nuclei $(Z+1, A)$ 
involved in our calculation are odd-Z nuclei. In these cases, the time reversal symmetry gets violated in the mean field models\cite{bha,block}. 
Therefore, to evaluate the  binding energy of odd-Z nuclei, we used the Pauli blocking prescription \cite{block} 
which restores the time-reversal symmetry.
In Fig .  \ref{Qbet} and Fig. \ref{tbet} our results for $Q_\beta$ and $T_{1/2}^{\beta}$ are displayed, respectively.  
For comparison, FRDM results are also given. One can see that RMF results almost follow 
the similar trend as that of FRDM results which confirms the applicability 
of the above formula\cite{fiset} in $Z=70-80$ mass region. Further, from Fig. \ref{tbet} we 
can conclude that the  decay of these nuclei mainly choose  $\beta$ decay as the prominent path. 	

\begin{table*}[htp!]
\caption{The RMF(NL3) results for binding energy BE
(MeV), Neutron, Proton, Charge, Root mean square radii(fm), quadrupole
deformation parameter $\beta_2$ and hexadecoupole parameter $\beta_4$ 
 for Yb isotopes compared with the Finite Range Droplet Model (FRDM) data ~\cite{moll97,moll95} and available
 experimental results ~\cite{audi12,raman01,angeli13}.} 
\renewcommand{\tabcolsep}{0.03 cm}
\renewcommand{\arraystretch}{0.9}
{
\begin{tabular}{|c|c|c|c|c|c|c|c|c|c|c|c|c|c|c|c|c|c|c|c|c|c|}
\hline
& \multicolumn{7}{ |c| }{RMF (NL3)} &\multicolumn{3}{ |c| }{FRDM}& 
\multicolumn{3}{ |c| }{Expt.}  \\
\cline{2-13}
\hline
Nucleus  & BE(R) & $r_{n}$&$r_{p}$&$ r_{ch}$ &$r_{rms}$& $\beta_{2}$ &$\beta_{4}$&
BE(F)& $\beta_{2}$ &$\beta_{4}$& BE(Ex)& $r_{ch}$ & $\beta_{2}$  \\
\hline
$^{150}$Yb		&	1198.1	&	5.05	& 	5.06	&	5.12	&	5.06	&	-0.15	&	-0.01	&	1194.6	&	-0.16	&	0.00	&		1194.6	&		&		\\
		
$^{152}$Yb 		&	1220.9	&	5.06	&	5.04	&	5.11	&	5.05	&	-0.00	&	0.00	& 	1218.6	& -0.00	&	0.00	&	1202.4	& 5.04	& 	\\
														
$^{154}$Yb		&	1240.4	&	5.11	&	5.06	&	5.12	&	5.09	&	-0.08	&	0.01 &  	1238.4	&	-0.01	&	0.00	&	1206.0	&	5.09	&   	\\

$^{156}$Yb 		&	1260.3	&	5.16	&	5.09	&	5.15	&	5.13	&	0.15	&	0.05	&	1257.4	&	0.13	&	0.03	&	1257.6	&	5.12	&  	\\

$^{158}$Yb		&	1279.1	&	5.20	&	5.11	&	5.18	&	5.16	&	0.19	&	0.06 &	1276.3	&	0.16	&	0.03	&	1276.5	&	5.15	&	0.19		\\

$^{160}$Yb		&	1297.0	&	5.25	&	5.14	&	5.2	&	5.20	&	0.22	&	0.06 &	1294.8	&	0.21	&	0.02	&	1294.8	&	5.18	&	0.23		\\
\end{tabular}\label{Yb} }

 
$^{198}$Yb		&	1516.7	&	5.82	&	5.35	&	5.41	&	5.66	&	0.07	&	0.02	&	1520.2	&	0.02	&	0.00	&	&		&	\\
										
$^{200}$Yb		&	1522.0 &	5.86	&	5.38	&	5.44	&	5.70	&	0.13	&	0.08	&	1523.4	&	0.06	&	0.04	&		&		&		\\
																	
$^{202}$Yb		&	1527.4	&	5.91	&	5.41	&	5.47	&	5.74	&	0.17	&	0.10	&  	1527.3	&	0.11	&	0.06	&		&		&	    \\
															
$^{204}$Yb		&	1532.1	&	5.95	&	5.43	&	5.49	&	5.78	&	0.19	&	0.09	& 	1531.2	&	0.14	&	0.07	&	&		&	  \\
																
$^{206}$Yb		&	1536.5	&	5.99	&	5.46	&	5.51	&	5.81	&	0.21	&	0.08	& 	1535.3	&	0.16	&	0.08	&		&		&	 \\
										
$^{208}$Yb 		&	1540.4	&	6.03	&	5.48	&	5.53	&	5.85	&	0.23	&	0.08	&	1539.2	&	0.18	&	0.09	&		&		&	  \\
$^{210}$Yb		&	1544.9	&	6.08	&	5.54	&	5.60	&	5.91	&	0.30	&	0.20	& 	1542.9	&	0.21	&	-0.08	&		&		&	   \\
$^{212}$Yb		&	1549.6	&	6.12	&	5.56	&	5.62	&	5.95	&	0.32	&	0.20	&1546.2	&	0.22	&	0.07	&		&		&	    \\
\hline
\hline
\end{tabular}\label{Yb} } 
\end{table*}

\begin{table*}
\caption{Same as Table-1 for Hf isotopes.} 
\renewcommand{\tabcolsep}{0.03cm}
\renewcommand{\arraystretch}{0.9}
{\begin{tabular}{|c|c|c|c|c|c|c|c|c|c|c|c|c|c|c|c|c|c|c|c|c|c|}
\hline
\hline
& \multicolumn{7}{ |c| }{RMF (NL3)} &\multicolumn{3}{ |c| }{FRDM}&
\multicolumn{3}{ |c| }{Expt.}  \\
\cline{2-13}
\hline
Nucleus& BE(R)&$r_{n}$&$r_{p}$&$ r_{ch}$ &$r_{rms}$& $\beta_{2}$ &$\beta_{4}$&
BE(F)& $\beta_{2}$ &$\beta_{4}$&
BE(Ex)& $r_{ch}$ & $\beta_{2}$ \\

\hline
$^{152}$Hf		&	1198.8	&	5.06	&	5.09	&	5.15	&	5.07	&	-0.13	&	-0.02	&	1194.4	&	-0.15	&	-0.05	&		&		&	\\
														
$^{154}$Hf 		&	1223.1	&	5.07	&	5.08	&	5.14	&	5.07	&	-0.00	&	0.00	&	1219.9	&	0.01	&	0.00	&		1219.4	&		&	    \\
											
$^{156}$Hf		&	1243.7	&	5.11	&	5.09	&	5.15	&	5.10	&	0.07	&	0.01	& 	1240.8	&	0.04	&	-0.01	&	1240.65	&		&	 \\
									
$^{158}$Hf		&	1264.5&	5.16	&	5.12	&	5.18	&	5.14	&	0.14	&	0.03	&		1260.8	&	0.11	&	0.02	&	1261.0	&		&      \\
											
$^{160}$Hf		&	1284.3	&	5.20	&	5.14	&	5.20	&	5.18	&	0.17	&	0.03	&		1280.8	&	0.15	&	0.02	&	1281.0	&		&	\\

$^{162}$Hf 		&	1303.3	&	5.25	&	5.16	&	5.22	&	5.21	&	0.20	&	0.03	&		1300.2	&	0.18	&	0.01	&	1300.4	&		&	0.16          	\\
\hline
\end{tabular} \label{Hf} }
\end{table*}

\begin{table*}
\caption{Same as Table-1 for W isotopes.} 
\renewcommand{\tabcolsep}{0.03cm}
\renewcommand{\arraystretch}{0.9}
{\begin{tabular}{|c|c|c|c|c|c|c|c|c|c|c|c|c|c|c|c|c|c|c|c|c|c|c|c}
\hline
\hline
& \multicolumn{7}{ |c| }{RMF (NL3)} &\multicolumn{3}{ |c| }{FRDM}&
\multicolumn{3}{ |c| }{Expt.}  \\
\cline{2-13}
\hline
Nucleus& BE(R)&$r_{n}$&$r_{p}$&$ r_{ch}$ &$r_{rms}$& $\beta_{2}$ &$\beta_{4}$&
BE(F)& $\beta_{2}$ &$\beta_{4}$&
BE(Ex)& $r_{ch}$ & $\beta_{2}$ \\
\hline
										
$^{154}$W		&	1198.0	&	5.06	&	5.12	&	5.18	&	5.09	&	-0.11	&	-0.02	&      	  1192.8	&	-0.12	&	-0.04	&		&		&	       	\\	
																
$^{156}$W		&	1224.1	&	5.08	&	5.11	&	5.17	&	5.09	&	-0.01	&	0.0 & 1219.4	&	0.01	&	0.00	&		&		&                \\
															
$^{158}$W		&	1245.6	&	5.12	&	5.12	&	5.19	&	5.12	&	-0.04 &	0.00	&		1241.7	&	0.02	&	0.00	&	1241.1	&		&	      \\
													
$^{160}$W		&	1267.1	&	5.16	&	5.15	&	5.21	&	5.16	&	0.11	&	0.01	&		1262.6	&	0.09	&	0.00	&	1262.9	&		&       \\
				                                                                                                          
$^{162}$W		&	1287.9	&	5.20	&	5.17	&	5.23	&	5.19	&	0.15	&	0.01	&		1283.5	&	0.13	&	0.02	&	1283.7	&		&	\\
													
$^{164}$W		&	1307.7	&	5.25	&	5.19	&	5.25	&	5.22	&	0.17	&	0.01	&	1303.9	&	0.16	&	0.01	&	1304.0	&		&		\\
\hline
\hline
\end{tabular}\label{W} }
\end{table*}

\begin{table*}
\caption{Same as Table-1 for Os isotopes.}
\renewcommand{\tabcolsep}{0.03cm}
\renewcommand{\arraystretch}{0.9}
{\begin{tabular}{|c|c|c|c|c|c|c|c|c|c|c|c|c|c|c|c|c|c|c|c|c|c|c|c|c|c|}
\hline
\hline
& \multicolumn{7}{ |c| }{RMF (NL3)} &\multicolumn{3}{ |c| }{FRDM}&
\multicolumn{3}{ |c| }{Expt.}  \\
\cline{2-13}

\hline
Nucleus& BE(R)&$r_{n}$&$r_{p}$&$ r_{ch}$ &$r_{rms}$& $\beta_{2}$ &$\beta_{4}$&
BE(F)& $\beta_{2}$ &$\beta_{4}$&
BE(Ex)&  $r_{ch}$ & $\beta_{2}$ \\

\hline
$^{156}$Os             &       1195.9  &       5.06    &       5.15    &       5.21     &       5.10     &       -0.08   & -0.02 &              &              &              &              &              &             \\
                                                                        
$^{158}$Os             &       1223.8  &       5.08    &       5.14    &       5.21    &       5.11    &       -0.00   &       0.00   &              &              &              &              &              &             \\
                                                                
$^{160}$Os             &       1246.4  &       5.12    &       5.16    &       5.22    &       5.14    &       -0.00   &       0.00   &         1241.0    &       0.01    &              &                   &              &           \\
                                                                 
$^{162}$Os             &       1287.9  &       5.20    &       5.17    &       5.23    &       5.19    &       0.15    &       0.00   &        1262.8  &       0.05    &       -0.01   &     1262.6  &              &            \\
$ ^{164}$Os             &       1307.7  &       5.25     &       5.19    &       5.25    &       5.22     &       0.17    &       0.00   &     1284.5  &       0.11    &       -0.00   &       1284.7  &              &        \\
                                                                 
$^{166}$Os             &       1310.3  &       5.24    &       5.21    &       5.27    &       5.23    &       0.13    &       -0.00   &        1305.8  &       0.13    &       -0.00   &       1305.8  &              &             \\
\hline
\hline
\end{tabular}\label{Os} }
\end{table*}

\begin{table*}
\caption{Same as Table-1 for Pt isotopes.} 
\renewcommand{\tabcolsep}{0.03cm}
\renewcommand{\arraystretch}{0.9}
{\begin{tabular}{|c|c|c|c|c|c|c|c|c|c|c|c|c|c|c|c|c|c|c|c|c|c|c|c|c|}
\hline
\hline
& \multicolumn{7}{ |c| }{RMF (NL3)} &\multicolumn{3}{ |c| }{FRDM}&
\multicolumn{3}{ |c| }{Expt.}  \\
\cline{2-13}
\hline
Nucleus& BE(R)&$r_{n}$&$r_{p}$&$ r_{ch}$ &$r_{rms}$& $\beta_{2}$ &$\beta_{4}$&
BE(F)& $\beta_{2}$ &$\beta_{4}$&
BE(Ex)& $r_{ch}$ & $\beta_{2}$  \\

\hline
$^{158}$Pt             &       1192.2  &       5.06    &       5.18    &       5.24    &       5.12    &       -0.00   &    0.00   &              &              &              &              &              &            \\
                                                                  
$^{160}$Pt             &       1222.1    &       5.09    &       5.18    &       5.24    &       5.13    &   -0.002           &  0.00  &             &              &              &              &              &           \\
                                                                
$^{162}$Pt             &       1246.0  &       5.13    &       5.19    &       5.25    &       5.16    &       -0.00   &       0.00   &                  &              &              &              &              &          \\
                                                                  
$^{164}$Pt             &       1268.2  &       5.16    &       5.20    &       5.27    &       5.18    &       -0.03   &       0.00   &             &              &              &              &              &            \\
                                                               
$^{166}$Pt             &       1290.3  &       5.20     &       5.22    &       5.28    &       5.21    &       0.07    &       0.00 &           1283.8  &       -0.07   &     -0.01       &       1283.7  &              &          \\
\hline
\hline
\end{tabular}\label{Pt} }
\end{table*}

\begin{table*}
\caption{Same as Table-1 for Hg isotopes.}
\renewcommand{\tabcolsep}{0.03cm}
\renewcommand{\arraystretch}{0.9}
{\begin{tabular}{|c|c|c|c|c|c|c|c|c|c|c|c|c|c|c|c|c|c|c|c|c|c|c|c|c|c|c|c|c|c|}
\hline
\hline
& \multicolumn{7}{ |c| }{RMF (NL3)} &\multicolumn{3}{ |c| }{FRDM}&
\multicolumn{3}{ |c| }{Expt.}  \\
\cline{2-13}
\hline
Nucleus& BE(R)&$r_{n}$&$r_{p}$&$ r_{ch}$ &$r_{rms}$& $\beta_{2}$ &$\beta_{4}$&
BE(F)& $\beta_{2}$ &$\beta_{4}$&
BE(Ex)& $r_{ch}$ & $\beta_{2}$ \\
\hline
$^{160}$Hg              &       1187.3  &       5.07    &       5.22    &       5.28    &       5.15    &    -0.00          &   0.00           &              &              &              &              &              &            \\
                                                     
$^{162}$Hg             &       1218.7  &       5.10     &       5.22     &       5.28    &       5.16    &   -0.00           &  0.00   &               &              &              &              &              &            \\
                                                                   
$^{164}$Hg             &       1243.9  &       5.13    &       5.22    &       5.29    &       5.18    &     -0.00       &       0.00   &        &              &              &              &              &         \\
                                                                   
$^{166}$Hg             &       1267.0  &       5.17    &       5.24    &       5.30    &       5.20     &       -0.00   &       0.00  &               &              &              &              &              &          \\
                                                               
$^{168}$Hg             &       1289.7  &       5.21    &       5.25    &       5.31     &       5.23    &       -0.01   &       0.00   &                     &              &              &              &              &        \\
\hline
\end{tabular}\label{Hg} }
\end{table*}

\section{Summary and Conclusion}\label{remark}
We have calculated quadrupole deformation, hexadecoupole deformation, two neutron separation energy and differential variation of two neutron separation energy
for some neutron-rich even-even nuclei in $Z=70-80$ region using RMF theory with pairing correlation from BCS approach (RMF-BCS).
The results from $S_{2n}$ and $dS_{2n}$
confirm the neutron shell closure at $N=82$ and $126$. 
As a further confirmatory test, the single particle energy levels 
for neutrons in isotopic chains are examined.
We observed large gaps at $N=82$ and $126$.
We have also calculated half lives for $\alpha$ and $\beta$ decay and seen that neutron-rich nuclei prefer  
$\beta$ decay rather than going to $\alpha$ decay mode. Further we concluded that the RMF-BCS theory provides a 
reasonably good description for all the considered isotopes.

\section{Acknowledgment}
The author S. Mahapatro  thanks Institute of Physics, Bhubaneswar, India for 
providing library and computer facilities for these calculations.

\end{document}